\documentstyle[preprint,aps]{revtex}
\begin{document}
\draft
\title{{Phase oscillations between two superconducting condensates in 
cuprate superconductors}}
\author{\Large P.V.Shevchenko$^{a}$, O.P.Sushkov$^{b}$}
\address
{School of Physics, The University of New South Wales,
Sydney 2052, Australia}
\maketitle
\begin{abstract}
Implications of the small Fermi surface 
 are discussed.
We demonstrate that superconductivity in this system can be described in terms
of two coupled condensates. 
The two condensates result in a collective excitation corresponding to 
the relative phase oscillation - a phason.  
We discuss the possibility of searching for this collective excitation in
the dynamic resistance of the SQUID.
\end{abstract}
\bigskip
Corresponding author: Pavel Shevchenko, School of Physics, The University of New South Wales, Sydney, 2052, Australia\\
ph.(612)9385 61 32 fax(612)9385 60 60 e-mail: pavel@newt.phys.unsw.edu.au
\bigskip\bigskip


\section{Introduction}
There is a lot of controversy about the shape of the Fermi surface in
cuprate superconductors.  In the early days it was believed that it is a
small Fermi surface of the doped Mott insulator \cite{Bed}. Later
many of the results from photoelectron spectroscopy (PES) have been
interpreted as favoring a large Fermi surface in agreement with
Lattinger's theorem \cite{Dessau}. On the other hand most recent PES
data\cite{Aebi,Chak,Wells,Mars} once more give indications of a small Fermi
surface for underdoped samples.
In the present paper we consider the scenario with a small
Fermi surface consisting of hole pockets around $(\pm \pi/2, \pm \pi/2)$,
see Fig.1a.
It is widely believed that the $t-J$ model
describes the main details of the doped Mott insulator. To fit the experimental
hole dispersion one needs to extend the model introducing additional hopping 
matrix elements $t^{\prime}$, $t^{\prime \prime}$ 
(see. e.g. Refs.\cite{Naz,Bala,Sus}), but basically it is the $t-J$ model. 
Superconducting pairing  
induced
by spin-wave exchange 
in the $t-J$ model
has been considered in the papers \cite{Flam,Bel}.
It was demonstrated \cite{Flam}  that there is an infinite set of
solutions for the superconducting gap. All the solutions have nodes
along the lines $(1,\pm 1)$, see Fig.1a. (It is very convenient to
use the magnetic Brillouin zone, but it can certainly  be mapped
to the full zone.) Using translation by the vector of the inverse magnetic lattice
the picture can be reduced to two hole pockets centered around
the points $(\pm \pi/2, \pi/2)$, see Fig.1a. The superconducting pairing is
the strongest between particles from the same pocket, and the lowest energy
solution for the superconducting gap has only one node line in each pocket.
Having this solution in a single pocket one can generate two solutions in
the whole Brillouin zone taking symmetric or antisymmetric combinations
between pockets. The symmetric combination corresponds to the d-wave
(Fig.1b), and the antisymmetric combination corresponds to the g-wave (Fig.1c)
pairing. 
We would like to note that the possibility of generating new solutions by taking
different combinations between pockets was first demonstrated
by Scalettar, Singh, and Zhang in the paper \cite{Scal}.
The energy splitting between the d- and g-wave solutions has been
investigated numerically in the Ref.\cite{Bel}.
The g-wave solution disappears and only the d-wave one survives
as soon as the hole dispersion is degenerate along the face of the magnetic
Brillouin zone. Actually in this situation there are no pockets
and one has a large open Fermi surface at an arbitrary small hole 
concentration.
However for small well separated pockets the d- and g-wave solutions are almost
degenerate. In pure $t-J$ or $t-t^{\prime}-t^{\prime \prime}-J$ model the
d-wave solution always has the lower energy. However if one extends the model
including the nearest sites hole-hole Coulomb repulsion the situation can be
inverted. The nearest sites repulsion does not influence the  g-wave pairing 
and substantially suppresses the d-wave pairing. So it is quite possible that 
the real ground state has the g-wave superconducting gap.
We would like to note that the g-wave ground state does not contradict the
existing experimental data on Josephson tunneling \cite{Harl}. The matter
is that in this case the tunneling current pumps 
g-wave into d-wave in a thin layer near the contact, and this gives the
interference picture very close to that for a pure d-wave ground 
state\cite{stat1}.
Anyway whatever the symmetry of the cuprate superconducting ground state 
(d-wave or g-wave), in the present paper we consider the scenario of the Fermi 
surface with separated hole pockets.
According to the 
microscopic picture 
described above
in this case we should 
consider simultaneously two coupled superconducting condensates which is 
equivalent to coexistence of the d- and g-wave pairings. A very strong evidence
in favor of multicomponent condensate has been the recent observation of
coherent Josephson response in microwave impedance of YBa$_2$Cu$_3$O$_{6.95}$ 
single crystals \cite{Zhai}.

The possibility of ``two-gap'' superconductivity has been suggested
for conventional superconductors a long time ago \cite{Walker}. 
Collective excitation 
corresponding to relative
phase oscillations between two condensates 
in such superconductors 
has been considered by
Legget \cite{Legget}. He also pointed out the difficulty of experimental
observation of this excitation.
This basic difficulty is that the excitation can be revealed only in 
oscillations of the relative density of the electrons in two bands,
and there is no external probe which is coupled directly to this quantity.

In the present work we  demonstrate existence of the phase  collective 
excitation in high $T_c$ cuprate superconductors and  calculate its 
dispersion in the long wavelenght limit. We  estimate the energy of this
excitation and compare it with the superconducting gap. We discuss also
the ways to search for the phase collective excitation.
Due to peculiar symmetry properties of the condensates (d-wave, g-wave)
the Josephson current through the tunnel contact between conventional 
and cuprate superconductor can pump one condensate to another
in some layer near the contact \cite{stat1}.
This gives the possibility of observing phase collective excitation in
the dynamic resistance of the contact or in the dynamic resistance
of SQUID.

\section{Ginzburg-Landau Lagrangian and free energy}

In the present paper we consider the scenario of the Fermi surface with well separated hole pockets in cuprate superconductors.
According to the 
microscopic picture 
described above 
we should consider
similtaneously the d- and g-wave pairings. Let us formulate an effective
Ginzburg-Landau theory describing this situation.
In the first approximation we can neglect the interaction between pockets
in k-space. Then a half of the holes belong to one pocket and the rest
belongs to the other pocket,
and we should introduce two macroscopic condensates corresponding to the
pockets,
$\Psi_1=\left|{\Psi_1}\right|\mbox {e}^{i\phi_1}$,
$\Psi_2=\left|{\Psi_2}\right|\mbox{e}^{i\phi_2}$,
where $\left|{\Psi_1}\right|=\left|{\Psi_2}\right|=\left|\Psi\right|=\sqrt{N_h}/2$,
$N_h$ is the number density of condensate holes.
The effective Lagrangian of the system in an external electric field $E$
is of the form (hereafter we set $\hbar=1$)
\begin{equation}
\label{L}
L=\sum_{n=1,2}\frac{i}{2}\left(\Psi_n^* \dot\Psi_n-\Psi_n \dot\Psi_n^* \right)-F
\end{equation}
where $F$ is Ginzburg-Landau free energy
\begin{equation}
\label{F}
F=\int\left\{\sum_{n=1,2}\left({1\over{2m^*}}|\nabla\Psi_n|^2
-a\left|\Psi_n\right|^2 +b\left|\Psi_n\right|^4
+2e\varphi\left[\left|\Psi_n\right|^2- N_h/4\right]\right)
+{{E^2}\over{8\pi}}\right\}dV+F_{\mbox{{\small int}}},
\end{equation}
with $\varphi$  a scalar potential, and
$F_{\mbox{{\small int}}}$  a small interaction between pockets.
Following Legget \cite{Legget} we use the simplest form of this interaction 
\begin{equation}
\label{Fint}
F_{\mbox{{\small int}}}=
\gamma \int \left(\Psi_1^*\Psi_2+\Psi_1\Psi_2^*\right)dV ,
\end{equation}
where $\gamma\ll a$ is a small parameter of the interaction. 
For the homogeneous case\\
 $F_{\mbox{{\small int}}}$ $\to$
$2\gamma V \left|\Psi_1\right|\left|\Psi_2\right|\cos(\phi_1-\phi_2)$,
where $V$ is the total volume.
The Hamiltonian corresponding to the Lagrangian (\ref{L}) is just
Ginzburg-Landau free energy (\ref{F}).
Note, that in the free energy we have neglected mass
anisotropy within the Fermi surface pocket. The anisotropy definitely
exists, and it is probably very important for nonlinear microwave response 
observed in the Ref. \cite{Zhai}. However the anisotropy does not
influence qualitatively the effects considered in the present work,
and therefore we neglect it for the sake of simplicity.
The equilibrium values of the order parameters are
\begin{equation}
\label{psi}
\left|\Psi_1\right|^2=\left|\Psi_2\right|^2=\left|\Psi\right|^2=
\frac{a+\left|\gamma\right|}{2b}\approx \frac{a}{2b}
.\end{equation}
The ground state phase difference $\Delta \phi= \phi_2 -\phi_1$
is determined by the sign of $\gamma$:
if $\gamma>0$, then $\Delta \phi=\pi$ (g-wave);
if $\gamma<0$, then $\Delta \phi=0$ (d-wave).
It is convenient also to introduce the d- and g-wave condensates
$\Psi_d=\Psi_1+\Psi_2=2\left|\Psi\right|\cos(\Delta\phi/2)e^{i\phi}$,
and
$\Psi_g=\Psi_1-\Psi_2=
2\left|\Psi\right|\sin(\Delta\phi/2)e^{i(\phi-\pi/2)}$,
where $\phi=(\phi_1+\phi_2)/2$.
So, the ground state has either {\bf d} or {\bf g}-wave symmetry:
if $\gamma>0$ then $\Psi_d=0, \Psi_g\neq 0$ and
if $\gamma<0$ then $\Psi_d\neq 0, \Psi_g=0$.

\section{Excitations}
Lagrange equations corresponding to (\ref{L}) are
\begin{eqnarray}
\label{LE}
i\frac{\partial \Psi_n}{\partial t}&=&
-\frac{\Delta}{2m^*}\Psi_n+2b\Psi_n
\left(\left|\Psi_n\right|^2-N_h/4\right)
+2e\varphi\Psi_n+|\gamma|\Psi_n + \gamma\Psi_{\bar n},\\
\Delta\varphi&=&-8\pi e\left(|\Psi_1|^2+|\Psi_2|^2-N_h/2\right),\nonumber
\end{eqnarray}
where ${\bar n}=2$ if n=1, and ${\bar n}=1$ if n=2.
Consider the case $\gamma<0$ which corresponds to the d-wave ground state. 
In this case $\Psi_n=\sqrt{N_h/4} + \delta \Psi_n$, where $\delta \Psi_n$
is the deviation from ground state value. Making a linear approximation in the
deviations, the eqs. (\ref{LE})  can be written as
\begin{eqnarray}
\label{LE1}
i\frac{\partial \delta\Psi_d}{\partial t}&=&
-\frac{\Delta}{2m^*}\delta \Psi_d+
\frac{bN_h}{2}(\delta\Psi_d+\delta\Psi_d^*)+2e\sqrt{N_h}\varphi,\nonumber\\
i\frac{\partial \delta\Psi_g}{\partial t}&=&
-\frac{\Delta}{2m^*}\delta\Psi_g+
\frac{bN_h}{2}(\delta\Psi_g+\delta\Psi_g^*)+2|\gamma|\delta\Psi_g,\\
\Delta\varphi&=&-4\pi e\sqrt{N_h}(\delta\Psi_d+\delta\Psi_d^*),\nonumber
\end{eqnarray}
where $\delta \Psi_d=\delta \Psi_1+\delta \Psi_2$, and
$\delta \Psi_g=\delta \Psi_1-\delta \Psi_2$.
One finds from eqs.(\ref{LE1}) that the d-wave oscillations
$\delta \Psi_d=A_d\exp(i{\bf kr}-i\omega t)+B_d^*\exp(-i{\bf kr}+i\omega t)$
correspond to the usual plasmon in charged Bose liquid with spectrum
\begin{equation}
\label{pl}
\omega^2_{\bf k}=\frac{8\pi e^2 N_h}{m^*}
+bN_h\frac{k^2}{2m^*}+\frac{k^4}{(2m^*)^2}
.\end{equation}
The phase oscillations which we are looking for are described by
\begin{equation}
\label{pg}
\delta \Psi_g=A_g e^{i{\bf kr}-i\omega t}+B_g^*e^{-i{\bf kr}+i\omega t}.
\end{equation}
From eq. (\ref{LE1}) one can easily find the dispersion of this excitation
\begin{equation}
\label{pha}
\omega^2_{\bf k}=2|\gamma| bN_h+4\gamma^2+
(bN_h+4|\gamma|)\frac{k^2}{2m^*}+
\frac{k^4}{(2m^*)^2}\approx 2|\gamma|bN_h+bN_h\frac{k^2}{2m^*},
\end{equation}
and the relation between $A_g$ and $B_g$
\begin{equation}
\label{ab}
B_g=-Z_{\bf k} A_g, \ \ \ \ Z_{\bf k}=1-2\frac{\omega_{{\bf k}}-k^2/(2m^*)-2|\gamma|}{bN_h}.
\end{equation}
To avoid misunderstanding we have to note that the phase oscillations
are always accompanied by relative density oscillations between the pockets.
Solution (\ref{pg}) represented in terms of 
phase and density variations looks like
\begin{eqnarray}
\label{pg1}
\phi_1&=&\phi_2=\phi_0 \sin({\bf kr}-\omega t),\\
{{\delta |\Psi_1|}\over{|\Psi_1|}}&=&-{{\delta |\Psi_2|}\over{|\Psi_2|}}
=\frac{\phi_0\omega_{\bf k}}{bN_h+k^2/(2m^*)+2|\gamma|} \cos({\bf kr}-\omega t),\nonumber
\end{eqnarray}
where $\phi_0$ is an amplitude of the phase oscillations.

The above consideration is relevant to the d-wave ground state ($\gamma < 0$),
however in the case of the g-wave ground state ($\gamma > 0$) all the results
are absolutely similar.

\section{Second quantization}
From the Lagrangian (\ref{L}), canonical momenta are
\begin{equation}
p_n={{\partial L}\over{\partial{\dot{\Psi_n}}}}=i\Psi_n^*,
\end{equation}
which gives $p_g={{i}\over{2}}\Psi_g^*$. With account of the relation
(\ref{ab}) the equation (\ref{pg}) can be rewritten in terms of the
Heisenberg creation operators of the phason $a_{\bf k}^{\dag}$:
\begin{equation}
\label{pgq}
\hat \Psi_g=\sum_{\bf k} Q_{\bf k}
\left(Z_{\bf k}^{-1/2}a_{\bf k}-Z_{\bf k}^{1/2}a_{\bf -k}^{\dag}\right)
e^{i{\bf kr}}.
\end{equation}
Quantization condition
\begin{equation}
\label{quant}
 [\hat p_g({\bf x}),\hat \Psi_g({\bf y})]={{i}\over{2}}
 [\hat \Psi_g^*({\bf x}),\hat \Psi_g({\bf y})]=-i\delta({\bf x-y}),
\end{equation}
together with standard commutation relations for creation and annihilation
operators
\begin{equation}
\label{aa}
 [a_{\bf k_1},a_{\bf k_2}]=0, \ \ \ \ 
 [a_{\bf k_1},a_{\bf k_2}^{\dag}]=\delta_{\bf k_1,k_2},
\end{equation}
gives the amplitude $Q_{\bf k}$ in equation (\ref{pgq})
\begin{equation}
\label{qk}
Q_{\bf k}=\sqrt{{2Z_k}\over{ V (1-Z_k^2)}}.
\end{equation}
After substitution of $\delta \Psi_1=-\delta \Psi_2={1\over{2}}\hat \Psi_g$
into Ginzburg-Landau free energy (\ref{F}) one finds the quantized Hamiltonian of
the system
\begin{equation}
\label{H}
\hat F=-{{bN_h^2}\over{8}}V
+\sum_{\bf k}\omega_{\bf k}\left(a_{\bf k}^{\dag}a_{\bf k}+\frac{1}{2}
\right).
\end{equation}
The first term here is the classical ground state energy, the second term
gives the spectrum of the phase excitation. 
We have considered the case of $\gamma <0$: d-wave ground state and g-wave 
phase excitations. If $\gamma >0$ then the  ground state has g-wave symmetry 
and the phase excitations correspond to the d-wave. However all the results,
spectra, etc. are not changed.

\section{Numerical estimations and discussion of the possibilities
for search of the phase excitation}
Due to eq. (\ref{pha}) the mimimal energy of the phase excitation is
$\omega_0=\omega_{k=0}\approx \sqrt{2|\gamma|bN_0}\approx 
2a\sqrt{|\gamma/a|}$. Let us demonstrate that this energy is much
smaller than the maximum of the superconducting gap. According to the
standard relation of Ginzburg-Landau theory, parameter $a$ is related to
the superconducting correlation length $\xi$: $a=1/(4m^*\xi^2)$. 
A typical value of this correlation length in cuprates is about 2-3
lattice spacing, and therefore
$a \sim 0.014eV/(m^*/m_e)$, where $m_e$ is an electron mass.
Taking rather arbitrarily  $|\gamma/a|\sim 1/10$ and $m^*/m_e \sim 7$ 
we find the frequency corresponding
to the phase excitation
\begin{equation}
\label{nu}
\hbar \omega_0 \sim 1mV, \ \ \ \ \nu_0 ={{\omega_0}\over{2\pi}} \sim 300GHz.
\end{equation}
The maximum of the superconducting gap on the Fermi surface $\Delta_{max}$
can be estimated by the standard BCS relation (see also discussion in 
Ref. \cite{Flam}): $\Delta_{max} \approx 2T_c \sim 200K$. This is 
probably the lowest possible estimation. This gives the following ratio of the
phase excitation energy to twice the superconducting gap
\begin{equation}
\label{rat}
{{\omega_0}\over{2\Delta_{max}}} \sim {{0.75}\over{m^*/m_e}}
\sqrt{\left|{{\gamma}\over{a}}\right|} \sim {1\over{30}},
\end{equation}
so it is really small. This  certainly does not mean that the
phason decay into particle-hole excitation is forbidden.
It is still allowed because the superconducting gap has nodes at the
Fermi surface. However due to the smallness of ratio (\ref{rat}) the
decay phase space is very small and therefore one should expect smallness 
of the decay width.

Relative phase oscillations in conventional superconductors were
predicted by Legget \cite{Legget}. He also pointed out that direct 
observation of these excitations is very complicated because there
is no external probe coupled to them.
Fortunately the situation in high-T$_c$ cuprate superconductors is
different. The phase oscillations can be excited in the tunneling contact
of a conventional superconductor with a cuprate. The matter is that
the supercurrent in such a contact under some conditions can pump the g-wave into 
d-wave and vice versa
in the layer of width $l_{\gamma}=\hbar/\sqrt{4|\gamma|m^*}$ near the
contact \cite{stat1}. Therefore time dependent supercurrent can excite
phase oscillations. So the idea is very simple: applying voltage $V$ to
the contact one induces oscillations of the supercurrent, and at 
$2eV=\omega > \omega_0$ absorption should sharply increase.
We repeat that this absorption arises only if the supercurrent drives
the relative phase difference in cuprate. The conditions under which it 
happens were investigated in our previous work\cite{stat1} and here
we present only conclusions of this work.
There are two possible scenarios: 1) the bulk ground 
state of the cuprate has pairing of d-wave symmetry, 2) the bulk ground state
of the cuprate has pairing of g-wave symmetry. 

1) Consider first the d-wave scenario. In this case the supercurrent in a single
tunnel contact of the cuprate with a conventional superconductor does not
drive the phase difference between the cuprate condensates, and one needs to 
consider the SQUID with $90^o$ cuprate
superconducting corner. There is no driving even in the SQUID 
if the sides of the corner are parallel to crystal axes $a$ and $b$.
So consider the corner rotated by some angle with respect to crystal
axes. In this case the supercurrent drives the relative phase at
zero magnetic flux in the SQUID ($\Phi=0$) and does not drive at $\Phi=0.5 \Phi_0$.
So in this situation the SQUID dynamic resistance depends on the magnetic flux
and at $\Phi=0$ one could observe the phase excitations.

2)In the case of the g-wave scenario the supercurrent drives the relative phase
even for a single tunnel contact and phase excitations can be seen in the
single contact dynamic resistance. Nevertheless it is interesting to
consider also the SQUID with 90$^o$ superconducting corner. If sides of the
corner are parallel to crystal axes $a$ and $b$, the current drives
the phase difference at an arbitrary magnetic flux, producing phase
excitations contributing to dynamic resistance.
If the corner is rotated by some angle the situation is more interesting:
There is no driving at $\Phi=0$ and maximum driving at $\Phi=0.5\Phi_0$.
This is exactly opposite to the d-wave case.

\section{Conclusions}

We have considered the scenario with the small Fermi surface 
consisting of hole pockets. The picture can be relevant
to underdoped cuprate superconductors. The small Fermi surface together
with mechanism of the magnetic pairing results in the possibility
of having both the d- and the g-wave pairing. Energy splitting
between these states is small. The ground state symmetry depends
on the interplay between the magnetic pairing and Coulomb repulsion.
We have demonstrated that these two condensates result in a collective
excitation corresponding to the relative phase oscillation - phason.
The energy of this collective excitation is of the order of 1mV which is much 
smaller than the maximum superconducting gap on the Fermi surface. The 
possibilities for searching for the phase excitation in the dynamic resistance
of a single tunnel junction and in the dynamic resistance of the SQUID are
discussed. These experiments allow also to determine the symmetry of the
ground state pairing.

\section{Acknowledgments}
We are very grateful to M.Kuchiev and D. van der Marel for stimulating 
discussions. This work was supported by a grant from the Australian Research 
Council.

\vspace{0.5cm}

FIGURE CAPTIONS

\vspace{0.5cm}

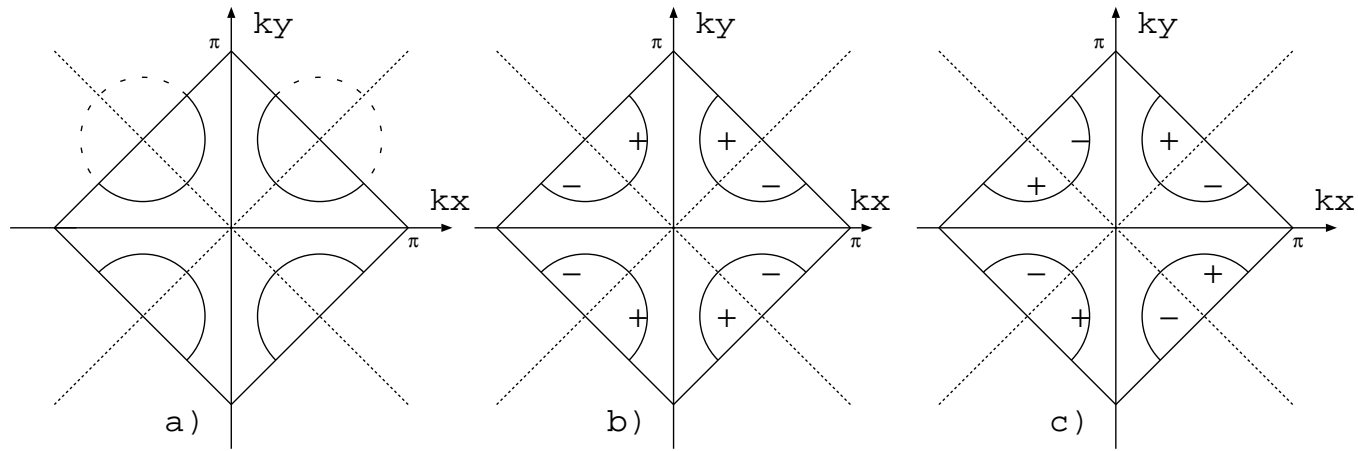
Fig.1.{\bf a}. Fermi surface in magnetic Brillouin zone which is equivalent to the two-pocket Fermi surface (dashed line). {\bf b}. Symmetry of the d-wave pairing in momentum space. {\bf c}. Symmetry of the g-wave pairing in momentum space.
\end{document}